\documentclass[prd,onecolumn,nofootinbib]{revtex4}
\usepackage{bm,amsmath,amssymb,graphicx}

\begin{document}

\noindent
Classical and Quantum Gravity {\bf 42} (6), 065017 (2025)\\

\title{Black holes in the expanding Universe}
\author{Nikodem Pop{\l}awski$^1$}
\altaffiliation{NPoplawski@newhaven.edu}

\affiliation{Department of Mathematics and Physics, University of New Haven, West Haven, CT, USA}

\begin{abstract}
The McVittie metric does not describe a physical black hole in an expanding Universe because the curvature scalar and pressure at its event horizon are infinite.
We show that extending this metric to an inhomogeneous scale factor, which depends on both the time and radial coordinate, removes those infinities by imposing at the horizon the constancy of the Hubble parameter and a particular constraint on the gradient of the scale factor.
We consider a special case of this metric, and show that the Hubble parameters at the event horizons of all centrally symmetric black holes are equal to the same constant $H_\textrm{hor}=(\Lambda/3)^{1/2}$.
Because of this equality and the equivalence to the Kottler metric near the horizon, black holes do not grow with the Universe expansion.
\end{abstract}
\maketitle

\noindent
{\bf 1. Introduction}\\
Black holes do exist \cite{blackhole}.
A centrally symmetric gravitational field of a nonrotating black hole in an asymptotically flat, static spacetime is described by the Schwarzschild metric \cite{Schw}.
The Universe is expanding \cite{expansion}.
On the largest scales, the Universe can be approximated as homogeneous and isotropic, and described by the Friedmann--Lema\^{i}tre--Robertson--Walker (FLRW) metric \cite{FLRW}.
A solution of the Einstein equations, describing the gravitational field of a centrally symmetric black hole or any massive body immersed in an expanding Universe, was proposed as the McVittie metric \cite{McVittie}.
It behaves like the Schwarzschild metric near the black hole and the FLRW metric far from the black hole.

The McVittie metric and its physical properties have been studied extensively \cite{Nolan}.
This solution, however, has a serious problem: the existence of a curvature singularity and an infinite pressure at the event horizon of a black hole \cite{singular}, unless the Hubble parameter of the expansion of the Universe is constant in time.
This constancy means that the McVittie solution is finite only in a Universe which is either static or expands exponentially.
Therefore, this metric is not a physical description of real black holes in the observed Universe.
It could be argued that the McVittie metric is regular because an event horizon in the system of reference of distant observers forms after an infinite time, and the Hubble parameter asymptotically tends to a constant value as the Universe expands \cite{Kaloper}.
For a complete analysis, however, one must use the comoving coordinates \cite{Tolman} to describe gravitational collapse of a spherical body and the subsequent formation of a black hole in the expanding Universe \cite{condensation}.

In this article, we extend the McVittie metric to an inhomogeneous scale factor, which depends on both the time and radial coordinate.
The resulting energy density is inhomogeneous, and the pressure is inhomogeneous and anisotropic, but far from the black hole they tend to homogeneous and isotropic functions of time, in accordance with the observed Universe on the large scale.
We show that imposing the constancy of the inhomogeneous Hubble parameter at an event horizon eliminates the infinity of the curvature scalar and pressure at the horizon.
We also show that the Hubble parameters at the event horizons of all centrally symmetric black holes in the absence of accreting matter are equal to the same constant, related to the cosmological constant.
Because of this equality and the consequential equivalence to the Kottler metric near the horizon, black holes do not grow with the Universe expansion.
The generalized McVittie metric with an inhomogeneous scale factor could also describe black holes accreting matter, embedded within galaxies and other large-scale structure instead of a region of the FLRW spacetime.\\

\noindent
{\bf 2. McVittie metric}\\
The McVittie metric for a black hole, embedded in the flat, expanding FLRW Universe, is given in the comoving, isotropic spherical coordinates $\rho,\theta,\phi$ (with $c=1$) by \cite{McVittie}
\begin{equation}
    ds^2=\Bigl(\frac{1-M}{1+M}\Bigr)^2 d\tau^2-(1+M)^4 a^2(d\rho^2+\rho^2 d\theta^2+\rho^2\sin^2\theta\,d\phi^2),\quad M(\tau,\rho)=\frac{Gm}{2a\rho},
    \label{metric}
\end{equation}
where $m$ is the mass of the black hole and $a(\tau)$ is the scale factor of the Universe, which is a function of the cosmic time $\tau$.
The Einstein equations for this metric, in the presence of an ideal fluid with energy density $\epsilon$ and pressure $p$ are:
\begin{equation}
    \kappa\epsilon=3H^2,\quad \kappa p=-3H^2-2\dot{H}\frac{1+M}{1-M},
    \label{fluid}
\end{equation}
where $H(\tau)=\dot{a}/a=(da/d\tau)/a=d(\ln a)/d\tau$ is the Hubble parameter, the dot denotes differentiation with respect to $\tau$, and $\kappa=8\pi G$.
The Ricci scalar is equal to $-12H^2-6\dot{H}(1+M)/(1-M)$.

For a stationary Universe, the metric (\ref{metric}) reduces to the Schwarzschild metric in the isotropic spherical coordinates \cite{Weyl,LL2}.
In the absence of a black hole ($m=0$), the metric (\ref{metric}) reduces to the flat FLRW metric.
The relations (\ref{fluid}) reduce to the Friedmann equations:
\begin{equation}
    \dot{a}^2=\frac{1}{3}\kappa\epsilon a^2,\quad \dot{a}^2+2a\ddot{a}=-\kappa p a^2.
    \label{reduce}
\end{equation}

\noindent
{\bf 3. Equivalency to Kottler metric}\\
A coordinate transformation into a new radial coordinate $r$ \cite{Robertson}:
\begin{equation}
    r=a(\tau)\rho(1+M)^2,
    \label{transf}
\end{equation}
transforms the metric (\ref{metric}) into
\begin{eqnarray}
    & & ds^2=\Bigl(1-\frac{2Gm}{r}-H^2 r^2\Bigr)d\tau^2+\frac{2Hr}{(1-2Gm/r)^{1/2}}d\tau\,dr-\frac{dr^2}{1-2Gm/r}-r^2 d\theta^2-r^2\sin^2\theta\,d\phi^2 \nonumber \\
    & & =f\Bigl(d\tau+\frac{Hr}{(1-2Gm/r)^{1/2}f}dr\Bigr)^2-\frac{1}{f}dr^2-r^2 d\theta^2-r^2\sin^2\theta\,d\phi^2,
    \label{nondiag}
\end{eqnarray}
where $f(\tau,r)=1-2Gm/r-H^2 r^2$.
All points having the same radial coordinate $r$ compose a sphere whose surface area is $4\pi r^2$.

If the Hubble parameter $H$ is constant, then a coordinate transformation into a new time coordinate $t$:
\begin{equation}
    t=\tau+H\int\frac{r}{(1-2Gm/r)^{1/2}f}dr,
    \label{time}
\end{equation}
transforms the nondiagonal metric (\ref{nondiag}) into the diagonal Kottler (Schwarzschild--de Sitter) metric \cite{Kottler}:
\begin{equation}
    ds^2=\Bigl(1-\frac{2Gm}{r}-\frac{1}{3}\Lambda r^2\Bigr)dt^2-\frac{dr^2}{1-2Gm/r-\Lambda r^2/3}-r^2 d\theta^2-r^2\sin^2\theta\,d\phi^2,
    \label{SdS}
\end{equation}
where $\Lambda=3H^2$ is a positive cosmological constant.
The cosmological constant is the simplest explanation of the observed accelerating expansion of the Universe \cite{dark}.
The Kottler metric describes a centrally symmetric black hole in a spacetime with a cosmological constant, reducing to the de Sitter metric \cite{deSitter} in the absence of the black hole and to the Schwarzschild metric in the absence of the cosmological constant.

Because of the equivalency to the Kottler metric, the McVittie metric with a constant $H$ does not provide any new information.
This case describes an exponential expansion of the Universe: $a\sim e^{H\tau}$, which is satisfied only for an empty Universe with a positive cosmological constant.
Therefore, it does not describe a black hole in the real Universe with the presence of matter.
Only the case with a time-varying Hubble parameter is of interest.\\

\noindent
{\bf 4. Event horizon}\\
The event horizon of a black hole represented by the McVittie metric (\ref{metric}) is given by $M=1$:
\begin{equation}
    \rho_\textrm{hor}=\frac{Gm}{2a}.
    \label{horizon}
\end{equation}
Because of the term $\dot{H}/(1-M)$ in (\ref{fluid}), the pressure at the horizon diverges at the horizon, unless $\dot{H}=0$ (constant $H$).
This metric has a curvature singularity at the horizon because the Ricci scalar also diverges there, unless $\dot{H}=0$.
The case with a constant $H$ is described by the Kottler metric (\ref{SdS}), for which the radius $r_\textrm{hor}$ of the event horizon of a black hole is constant and given by a cubic equation $1-2Gm/r_\textrm{hor}-\Lambda r^2_\textrm{hor}/3=0$.
However, it does not describe a black hole in the real Universe, which has a time-varying Hubble parameter.
Therefore, the McVittie metric does not describe a physical black hole embedded in the FLRW spacetime because the curvature and pressure must be finite \cite{Nolan,singular}.

To remove the curvature and pressure singularities at the event horizon, one must impose there $\dot{H}=0$.
However, astronomical observations show that the Hubble parameter is changing in time, so $\dot{H}\neq 0$.
Consequently, the scale factor at the event horizon must be different from the scale factor of the Universe as a whole.
The scale factor must depend not only on the time $\tau$, but also on the radial coordinate $\rho$.
Near a black hole, the scale factor is inhomogeneous.
Far from a black hole, where $M\approx 0$, the scale factor and the Hubble parameter should asymptotically tend to functions of the time only, in accordance with the observed homogeneity and isotropy of the Universe on the large scale (the Cosmological Principle).\\

\noindent
{\bf 5. Tolman metric}\\
We consider the generalized metric (\ref{metric}) with the scale factor depending on both the time and radial coordinate.
To determine the resulting field equations, we use the Tolman metric, which describes the geometry of a centrally symmetric gravitational field in spacetime filled with an ideal fluid \cite{Tolman,LL2}:
\begin{equation}
    ds^2=e^{\nu(\tau,R)}d\tau^2-e^{\lambda(\tau,R)}dR^2-e^{\mu(\tau,R)}(d\theta^2+\mbox{sin}^2\theta\,d\phi^2),
    \label{grav1}
\end{equation}
where $\nu$, $\lambda$, and $\mu$ are functions of a time coordinate $\tau$ and a radial coordinate $R$.
Applying coordinate transformations $\tau\rightarrow \tilde{\tau}(\tau)$ and $R\rightarrow \tilde{R}(R)$ does not change the form of the metric (\ref{grav1}).
The components of the Einstein tensor corresponding to (\ref{grav1}) that do not vanish identically are \cite{Tolman,LL2}:
\begin{eqnarray}
    & & G_0^0=-e^{-\lambda}\Bigl(\mu''+\frac{3}{4}\mu'^2-\frac{1}{2}\mu'\lambda'\Bigr)+\frac{1}{2}e^{-\nu}\Bigl(\dot{\lambda}\dot{\mu}+\frac{1}{2}\dot{\mu}^2\Bigr)+e^{-\mu}, \nonumber \\
    & & G_1^1=-\frac{1}{2}e^{-\lambda}\Bigl(\frac{1}{2}\mu'^2+\mu'\nu'\Bigr)+e^{-\nu}\Bigl(\ddot{\mu}-\frac{1}{2}\dot{\mu}\dot{\nu}+\frac{3}{4}\dot{\mu}^2\Bigr)+e^{-\mu}, \nonumber \\
    & & G_2^2=G_3^3=-\frac{1}{4}e^{-\lambda}(2\nu''+\nu'^2+2\mu''+\mu'^2-\mu'\lambda'-\nu'\lambda'+\mu'\nu')-\frac{1}{4}e^{-\nu}(\dot{\lambda}\dot{\nu}+\dot{\mu}\dot{\nu}-\dot{\lambda}\dot{\mu}-2\ddot{\lambda}-\dot{\lambda}^2-2\ddot{\mu}-\dot{\mu}^2), \nonumber \\
    & & G_0^1=\frac{1}{2}e^{-\lambda}(2\dot{\mu}'+\dot{\mu}\mu'-\dot{\lambda}\mu'-\dot{\mu}\nu'),
    \label{grav2}
\end{eqnarray}
where the prime denotes differentiation with respect to $R$.

In the cosmologically comoving frame of reference, the nonzero components of the energy--momentum tensor for the fluid are: $T^0_0=\epsilon$, $T^1_1=-p_\textrm{r}$, $T^2_2=T^3_3=-p_\textrm{t}$, and $T^1_0=S$, where $p_\textrm{r}$ is the radial pressure of the fluid, $p_\textrm{t}$ is the tangential pressure, and $S$ is the energy flux density.
The pressure is generally anisotropic, but the spherical symmetry is preserved.
The flux density $S$ generally does not vanish, because the fluid may move radially relative to the cosmologically comoving coordinates during accretion.
The Einstein field equations $G^i_k=\kappa T^i_k$ in this frame are:
\begin{equation}
    G_0^0=\kappa\epsilon,\quad G_1^1=-\kappa p_\textrm{r},\quad G_2^2=G_3^3=-\kappa p_\textrm{t},\quad G_0^1=\kappa S.
    \label{grav3}
\end{equation}
The covariant conservation of the energy--momentum tensor gives
\begin{eqnarray}
    & & \dot{\epsilon}+S'+\Bigl(\frac{1}{2}\dot{\lambda}+\dot{\mu}\Bigr)\epsilon+\Bigl(\frac{1}{2}(\nu'+\lambda')+\mu'\Bigr)S+\frac{1}{2}\dot{\lambda}p_\textrm{r}+\dot{\mu}p_\textrm{t}=0, \nonumber \\
    & & e^{\lambda-\nu}\dot{S}+p'_\textrm{r}+\Bigl(\frac{1}{2}(3\dot{\lambda}-\dot{\nu})+\dot{\mu}\Bigr)e^{\lambda-\nu}S+\frac{1}{2}\nu'(\epsilon+p_\textrm{r})+\mu'(p_\textrm{r}-p_\textrm{t})=0.
    \label{grav4}
\end{eqnarray}

In the isotropic spherical coordinates, the functions $\lambda$ and $\mu$ are related to each other by a condition $e^\mu=R^2 e^\lambda$.
The Tolman metric (\ref{grav1}) becomes
\begin{equation}
    ds^2=e^{\nu(\tau,R)}d\tau^2-e^{\lambda(\tau,R)}(dR^2+R^2 d\theta^2+R^2\mbox{sin}^2\theta\,d\phi^2).
    \label{grav5}
\end{equation}
Accordingly, $\mu=2\ln R+\lambda$, $\mu'=2/R+\lambda'$, $\mu''=-2/R^2+\lambda''$, $\dot{\mu}=\dot{\lambda}$, and $\ddot{\mu}=\ddot{\lambda}$. 
The Einstein equations (\ref{grav3}) for the components of the Einstein tensor (\ref{grav2}) reduce to
\begin{eqnarray}
    & & \kappa\epsilon=-e^{-\lambda}\Bigl(\lambda''+\frac{1}{4}\lambda'^2+\frac{2\lambda'}{r}\Bigr)+\frac{3}{4}e^{-\nu}\dot{\lambda}^2, \nonumber \\
    & & \kappa p_\textrm{r}=\frac{1}{2}e^{-\lambda}\Bigl(\frac{1}{2}\lambda'^2+\frac{2\lambda'}{r}+\lambda'\nu'+\frac{2\nu'}{r}\Bigr)-e^{-\nu}\Bigl(\ddot{\lambda}-\frac{1}{2}\dot{\lambda}\dot{\nu}+\frac{3}{4}\dot{\lambda}^2\Bigr), \nonumber \\
    & & \kappa p_\textrm{t}=\frac{1}{4}e^{-\lambda}\Bigl(2\nu''+\nu'^2+2\lambda''+\frac{2\lambda'}{r}+\frac{2\nu'}{r}\Bigr)+\frac{1}{4}e^{-\nu}(2\dot{\lambda}\dot{\nu}-3\dot{\lambda}^2-4\ddot{\lambda}), \nonumber \\
    & & \kappa S=\frac{1}{2}e^{-\lambda}(2\dot{\lambda}'-\dot{\lambda}\nu').
    \label{grav6}
\end{eqnarray}
The conservation relations (\ref{grav4}) reduce to
\begin{eqnarray}
    & & \dot{\epsilon}+S'+\frac{3}{2}\dot{\lambda}\epsilon+\Bigl(\frac{1}{2}\nu'+\frac{3}{2}\lambda'\Bigr)S+\dot{\lambda}\Bigl(\frac{1}{2}p_\textrm{r}+p_\textrm{t}\Bigr)=0, \nonumber \\
    & & e^{\lambda-\nu}\dot{S}+p'_\textrm{r}+\frac{1}{2}(5\dot{\lambda}-\dot{\nu})e^{\lambda-\nu}S+\frac{1}{2}\nu'(\epsilon+p_\textrm{r})+\lambda'(p_\textrm{r}-p_\textrm{t})=0.
    \label{grav7}
\end{eqnarray}

\noindent
{\bf 6. Field equations for generalized McVittie metric}\\
The metric (\ref{metric}) with an inhomogeneous scale factor $a(\tau,\rho)$:
\begin{equation}
    ds^2=\Bigl(\frac{1-M}{1+M}\Bigr)^2 d\tau^2-(1+M)^4 a^2(\tau,\rho)(d\rho^2+\rho^2 d\theta^2+\rho^2\sin^2\theta\,d\phi^2),\quad M(\tau,\rho)=\frac{Gm}{2a(\tau,\rho)\rho},
    \label{general}
\end{equation}
is a special case of the Tolman metric (\ref{grav5}) for
\begin{equation}
    R=\rho,\quad e^\nu=\Bigl(\frac{1-M}{1+M}\Bigr)^2,\quad e^\lambda=(1+M)^4 a^2.
\end{equation}
Because the scale factor depends on both $\tau$ and $\rho$, to define the Hubble parameter as a partial derivative $\partial(\ln a)/\partial\tau$, it must be specified which radial coordinate is regarded as constant; the partial derivatives $[\partial(\ln a)/\partial\tau]_\rho$ and $[\partial(\ln a)/\partial\tau]_r$ with $r=a\rho(1+M)^2$ (\ref{transf}) are different.
We define the Hubble parameter (and the dot in general) as the partial derivative with respect to time at a constant isotropic radial coordinate $\rho$:
\begin{equation}
    H(\tau,\rho)=\Bigl(\frac{\dot{a}}{a}\Bigr)_\rho=\frac{1}{a}\Bigl(\frac{\partial a}{\partial\tau}\Bigr)_\rho=\Bigl(\frac{\partial(\ln a)}{\partial\tau}\Bigr)_\rho.
\end{equation}
It is inhomogeneous like the scale factor.
We also define
\begin{equation}
    F(\tau,\rho)=\Bigl(\frac{a'}{a}\Bigr)_\tau=\frac{1}{a}\Bigl(\frac{\partial a}{\partial\rho}\Bigr)_\tau=\Bigl(\frac{\partial(\ln a)}{\partial\rho}\Bigr)_\tau,
\end{equation}
which characterizes the gradient of the scale factor.
Similarly, $\dot{H}=(\partial H/\partial\tau)_\rho$ and $H'=(\partial H/\partial\rho)_\tau$.
The prime denotes partial differentiation with respect to the isotropic radial coordinate $\rho$ at a constant time.

The following relations are satisfied:
\begin{eqnarray}
    & & M'=-M\Bigl(\frac{1}{\rho}+F\Bigr),\quad \dot{M}=-MH,\quad \nu'=\frac{4M}{1-M^2}\Bigl(\frac{1}{\rho}+F\Bigr),\quad \lambda'=-\frac{4M}{\rho(1+M)}+2F\frac{1-M}{1+M}, \nonumber \\
    & & \nu''=-\frac{4M(1+M^2)}{(1-M^2)^2}\Bigl(\frac{1}{\rho}+F\Bigr)^2+\frac{4M}{1-M^2}\Bigl(-\frac{1}{\rho^2}+F'\Bigr),\quad \dot{\nu}=\frac{4MH}{1-M^2},\quad \dot{\lambda}=2H\frac{1-M}{1+M}, \nonumber \\
    & & \lambda''=\frac{4M}{(1+M)^2}\Bigl(\frac{1}{\rho}+F\Bigr)^2+\frac{4M}{\rho^2(1+M)}+2F'\frac{1-M}{1+M},\quad \ddot{\lambda}=2\dot{H}\frac{1-M}{1+M}+\frac{4MH^2}{(1+M)^2},\nonumber \\
    & & \dot{\lambda}'=2H'\frac{1-M}{1+M}+\frac{4MH}{(1+M)^2}\Bigl(\frac{1}{\rho}+F\Bigr).
    \label{relations}
\end{eqnarray}
Consequently, the Einstein equations (\ref{grav6}) give
\begin{eqnarray}
    & & \kappa\epsilon=3H^2-\frac{1}{a^2(1+M)^5}\Bigl(\frac{4F}{\rho}+F^2(1+M)+2F'(1-M)\Bigr),
    \label{field1} \\
    & & \kappa p_\textrm{r}=-3H^2-2\dot{H}\frac{1+M}{1-M}+\frac{1}{a^2(1+M)^4}\Bigl(\frac{2F}{\rho}+F^2\Bigr), \label{field2} \\
    & & \kappa p_\textrm{t}=-3H^2-2\dot{H}\frac{1+M}{1-M}+\frac{1}{a^2(1+M)^5}\Bigl[\frac{1+M^2}{1-M}\Bigl(\frac{F}{\rho}+F'\Bigr)-\frac{2MF^2}{1+M}\Bigr],
    \label{field3} \\
    & & \kappa S=\frac{2H'(1-M)}{a^2(1+M)^5}.
    \label{field4}
\end{eqnarray}
The Ricci scalar is equal to $\kappa(p_\textrm{r}+2p_\textrm{t}-\epsilon)$.
These equations give the space and time dependence of the energy density, radial pressure, tangential pressure, and energy flux density near a black hole, described by the inhomogeneous McVittie metric.
They generalize the relations (\ref{fluid}) to an inhomogeneous scale factor $a(\tau,\rho)$.
The radial and tangential pressures are different; the pressure is anisotropic.
The energy and both pressures are inhomogeneous; they depend on the radial coordinate $\rho$.\\

\noindent
{\bf 7. Regularity at an event horizon}\\
A singular behavior of the metric (\ref{general}) at the event horizon of a black hole, where $M=1$, is a result of a singular behavior of $\nu'$, $\nu''$, $\dot{\nu}$, and $e^{-\nu}$.
To ensure the regularity of the pressure and curvature for this metric at the horizon, equations (\ref{field2}) and (\ref{field3}) give
\begin{equation}
    \dot{H}_\textrm{hor}=0.
    \label{first}
\end{equation}
Furthermore, to ensure the regularity at the horizon, equation (\ref{field3}) gives $(F/\rho+F')_\textrm{hor}=0$, which constrains the gradient of the scale factor.
Let us consider a special case, in which the above relation is satisfied for an arbitrary $\rho$:
\begin{equation}
    \frac{F}{\rho}+F'=0,\quad F=\frac{A(\tau)}{\rho},
    \label{second}
\end{equation}
where $A(\tau)$ is some function of the time coordinate.
Consequently, $(\ln a)'=A(\tau)/\rho$, which integrated with respect to $\rho$ yields $\ln a=A(\tau)\ln\rho+\ln B(\tau)$, where $B(\tau)$ is some function of the time coordinate.
The scale factor is therefore given by
\begin{equation}
    a(\tau,\rho)=B(\tau)\rho^{A(\tau)}.
    \label{scale}
\end{equation}
Accordingly, $\dot{a}=\dot{B}\rho^A+B\dot{A}\rho^A\ln\rho$, which gives
\begin{equation}
    H=\frac{\dot{B}}{B}+\dot{A}\ln\rho=(\ln B)^\cdot+\dot{A}\ln\rho,\quad H'=\frac{\dot{A}}{\rho},\quad \dot{H}=(\ln B)^{\cdot\cdot}+\ddot{A}\ln\rho,\quad \dot{H}'=\frac{\ddot{A}}{\rho}.
    \label{Hubble}
\end{equation}

At the horizon of a black hole, the relations (\ref{horizon}) and (\ref{scale}) give
\begin{equation}
    \frac{Gm}{2\rho}=B\rho^A,\quad \rho=\Bigl(\frac{Gm}{2B}\Bigr)^{1/(1+A)}.
    \label{horizon}
\end{equation}
The regularity condition (\ref{first}) with $\dot{H}$ given by (\ref{Hubble}) gives therefore
\begin{equation}
    (\ln B)^{\cdot\cdot}+\frac{\ddot{A}}{1+A}\ln\Bigl(\frac{Gm}{2B}\Bigr)=0.
    \label{regularity}
\end{equation}
This equation relates the functions $A(\tau)$ and $B(\tau)$ in the scale factor (\ref{scale}).
The metric (\ref{general}) with this scale factor is regular at the horizon if the limit $\lim_{M\to 1}\dot{H}/(1-M)$ of the term $\dot{H}/(1-M)$ in equations (\ref{field2}) and (\ref{field3}) is finite.
This term is indeterminate, of the form $0/0$.
Resolving the indeterminacy by L'H\^opital's rule and using (\ref{relations}), (\ref{second}), (\ref{Hubble}), and (\ref{regularity}), we obtain
\begin{equation}
    \lim_{M\to 1}\frac{\dot{H}}{1-M}=-\lim_{M\to 1}\frac{\dot{H}'}{M'}=\lim_{M\to 1}\frac{\ddot{A}/\rho}{M(1/\rho+F)}=\lim_{M\to 1}\frac{\ddot{A}}{1+\rho F}=\lim_{M\to 1}\frac{\ddot{A}}{1+A}=-\frac{(\ln B)^{\cdot\cdot}}{\ln(Gm/2B)}.
    \label{limit}
\end{equation}

In the scale factor (\ref{scale}), the function $B(\tau)$ describes the background scale factor, which exists in the absence of a black hole.
Therefore, $(\ln B)^{\cdot\cdot}$ is finite.
Also, $B\gg Gm$ (the Universe is much larger than the Schwarzschild radius of a black hole).
Consequently, the limit (\ref{limit}) is finite and the metric (\ref{general}) is regular at the horizon.

The function $A(\tau)$ describes the variation of the scale factor (\ref{scale}) with the radial coordinate $\rho$.
It must satisfy
\begin{equation}
    |A(\tau)|\ll 1,
    \label{small}
\end{equation}
so that the scale factor is approximately homogeneous, and $\lim_{\tau\to\infty}A=0$, so the Universe tends to a homogeneous form: $a=B(\tau)$.
The function (\ref{scale}) would be exact for all values of radial coordinate $\rho$ if the black hole considered were the only body in the Universe.
Therefore, its application is limited to the values of $\rho$, which are not too large, so the effects of other bodies on the metric can be averaged as the background scale factor.
In the absence of a black hole, $m=0$ gives $A=0$, in accordance with (\ref{regularity}), and the scale factor (\ref{scale}) is equal to its background form $a=B(\tau)$.

The energy density, radial pressure, tangential pressure, and energy flux density are determined by the field equations (\ref{field1}), (\ref{field2}), (\ref{field3}), and (\ref{field4}), with $F$ given by (\ref{second}) and $H$ given by (\ref{Hubble}).
They depend on the two functions $A(\tau)$ and $B(\tau)$, related to each other by (\ref{regularity}).
They are finite at the horizon because the functions $F$ and $H$, their derivatives $F'$ and $H'$, and the limit (\ref{limit}) are finite at the horizon.
The energy flux density in (\ref{field4}) at the horizon is zero, which is consistent with the infinite gravitational time dilation of accretion of matter, represented by the background energy density and pressure, onto a black hole described by the metric (\ref{general}).
Because of this dilation, the mass $m$ in the metric (\ref{general}) remains constant.
For a more complete analysis, one must investigate whether the energy density and pressure determined by the field equations (\ref{field1}), (\ref{field2}), (\ref{field3}), and (\ref{field4}) can represent the types of matter and fields existing in Nature.

Far from a black hole, $F\to 0$, $F'\to 0$, and $H'\to 0$, so the field equations (\ref{field1}), (\ref{field2}), (\ref{field3}), and (\ref{field4}) reduce to equations (\ref{fluid}) with $M\to 0$ (equivalent to the Friedmann equations (\ref{reduce}) for $A\to 0$) for a nearly homogeneous and isotropic fluid with $p_\textrm{t}=p_\textrm{r}$ and without energy flux, $S=0$.
The condition (\ref{small}) guarantees that the metric (\ref{general}) with the condition (\ref{second}) can describe a black hole embedded in the Universe, which on the large scale is nearly homogeneous and isotropic.

The above analysis describes the special case of the metric, represented by the condition (\ref{second}) on the gradient of the scale factor.
In a more general solution, which can include a black hole accreting matter, for which the mass is a function $m(\tau,\rho)$, that condition must be satisfied only at the horizon (to ensure the regularity).
More generally, the scale factor has the form (\ref{scale}) only near a black hole.
Far from the black hole, the scale factor should not vary with $\rho$, but instead tend to $B(\tau)$ as $\rho\to\infty$, in accordance with the homogeneity and isotropy of the Universe on the large scale.\\

\noindent
{\bf 8. Constancy of Hubble parameter at an event horizon}\\
The condition (\ref{first}) means that the Hubble parameter at the event horizon of a black hole is constant in time.
Using (\ref{Hubble}), (\ref{horizon}), and (\ref{regularity}), it is equal to
\begin{equation}
    H_\textrm{hor}=\lim_{M\to 1}[(\ln B)^\cdot+\dot{A}\ln\rho]=(\ln B)^\cdot+\frac{\dot{A}}{1+A}\ln\Bigl(\frac{Gm}{2B}\Bigr)=(\ln B)^\cdot-\frac{\dot{A}}{\ddot{A}}(\ln B)^{\cdot\cdot}.
    \label{constant}
\end{equation}
This constant value is equal to its value at $\tau\to\infty$, given by the asymptotic values $A\to 0$ and $B\to e^{H_0 \tau}$, where $H_0=(\Lambda/3)^{1/2}$ (for $\Lambda\ge 0$), giving $\ln B\to H_0\tau$, $(\ln B)^\cdot\to H_0$, and $(\ln B)^{\cdot\cdot}\to 0$.
Consequently, the Hubble parameter at an event horizon is equal to
\begin{equation}
    H_\textrm{hor}=\Bigl(\frac{\Lambda}{3}\Bigr)^{1/2},
    \label{cosmological}
\end{equation}
independently of the mass of a black hole.

Because the value (\ref{cosmological}) is a limit (\ref{constant}), it is valid not only in the special case of the metric, represented by the condition (\ref{second}), but also in a more general solution, in which that condition is satisfied only at the horizon (to ensure the regularity).
Therefore, the Hubble parameters at the event horizons of all centrally symmetric black holes are equal.

This result equates two boundary values of $H$ at different ends of the interval: an event horizon $\rho=\rho_\textrm{hor}$ and a future infinity $\tau\to\infty$.
It is consistent with the actual formation of an event horizon after an infinite cosmic time $\tau$ because of gravitational time dilation \cite{Tolman,LL2}.
As $\tau\to\infty$, the Friedmann equations (\ref{reduce}) in the presence of a positive cosmological constant give $H\to(\Lambda/3)^{1/2}$.
Without the cosmological constant, $H\to 0$.
In both cases, $\dot{H}\to 0$ as $\tau\to\infty$, so the regularization of an event horizon occurs in the same time limit as its formation \cite{Kaloper}.
A more complete consideration should use the comoving coordinates \cite{Tolman} to describe gravitational collapse, forming a black hole, in an expanding Universe \cite{condensation}.\\

\noindent
{\bf 9. Effect of Universe expansion on an event horizon}\\
At the event horizon of a black hole, described by the metric (\ref{general}), the Hubble parameter is constant and equal to (\ref{cosmological}).
Consequently, in the infinitesimal vicinity of such a black hole, where the scale factor does not vary significantly with $\rho$, the transformations (\ref{transf}) and (\ref{time}) turn the metric (\ref{general}) into the Kottler metric (\ref{SdS}).
Therefore, a black hole does not grow with the expansion of the Universe; it can grow only as a result of accretion.
Points in the space near the horizon move away from it, as the space expands, but the size of the horizon $r_\textrm{hor}$ remains constant in time.\\

\noindent
{\bf 10. Conclusion}\\
The McVittie metric does not describe a physical black hole in an expanding Universe because the curvature scalar and pressure at the event horizon are infinite.
We extended this metric to a metric with an inhomogeneous scale factor, which depends on both the time and radial coordinate (\ref{general}).
The resulting energy density is inhomogeneous, and the pressure is inhomogeneous and anisotropic.
We showed that the inhomogeneity of the scale factor eliminates those infinities by imposing the conditions (\ref{first}) and (\ref{second}) at the horizon.
We considered a special case of the metric, for which the second condition is satisfied also outside the horizon.
We found the corresponding form of the scale factor (\ref{scale}), satisfying the regularity condition (\ref{regularity}), and a condition under which the energy density and pressure far from the black hole tend to nearly homogeneous and isotropic functions of time, in accordance with the observed Universe on the large scale.

We showed that the Hubble parameter at the event horizon of every centrally symmetric black hole is equal to the same constant, determined by the cosmological constant (including zero) in the relation (\ref{cosmological}).
We also showed that, because of the resulting equivalence of the generalized McVittie metric and the Kottler metric, black holes do not grow with the Universe expansion.
This result is contrary to a hypothesis in \cite{growth} that black holes may be growing as the Universe expands, which aimed to explain how the enormous supermassive black holes at the centers of most galaxies have formed, and agrees with the results of \cite{Rudeep}.
Black holes can grow only if they accrete matter or merge with other black holes.

The results of this work are also valid in the Einstein--Cartan theory of gravity \cite{EC}, which removes the symmetry constraint on the affine connection and relates its antisymmetric part, the torsion tensor \cite{Niko}, to the intrinsic angular momentum (spin) of the matter composed of fermions, described by the Dirac equation.
In the presence of torsion, which manifests at extremely high densities as a repulsive force, gravitational singularities in black holes may be eliminated, and the singular big bang may be replaced by a regular big bounce \cite{universe}.
In vacuum, torsion vanishes and this theory reduces to general relativity, passing all its observational tests and giving the regularizing conditions (\ref{first}) and (\ref{second}) at an event horizon.

I am grateful to Francisco Guedes and my Parents, Bo\.{z}enna Pop{\l}awska and Janusz Pop{\l}awski, for supporting this work.
I thank the reviewers for their helpful comments and suggestions, which have strengthened the presented analysis.

\end{document}